# Re-assembly and test of a COMB dipole magnet with STAR® wires

V. V. Kashikhin, S. Cohan, J. DiMarco, O. Kiemschies, S. Krave, V. Lombardo, V. Marinozzi, D. Orris,
S. Stoynev, D. Turrioni, A. K. Chavda, U. Sambangi, S. Korupolu, J. Peram, A. Arjun, C. Goel, J. Sai Sandra,
V. Yerraguravagari, R. Schmidt, V. Selvamanickam, G. Majkic, E. Galstyan, N. Mai and K. Selvamanickam

*Abstract*—Rare-Earth Barium Copper Oxide (REBCO) coated conductors are an attractive option for application in high field accelerator magnets due to their high critical field and the convenience of fabrication without heat treatment compared to some other superconductors.

A small REBCO accelerator magnet was previously fabricated and tested in liquid nitrogen, demonstrating over 90% critical current retention in the coils. This paper describes the magnet re-assembly with a different support structure and its test in liquid helium at 1.8-4.5 K. The magnet quench history along with the instrumentation data is presented and discussed.

*Index Terms*—Accelerator magnets, coils, high-temperature superconductors, superconducting magnets, yttrium barium copper oxide.

## I. INTRODUCTION

FERMILAB is involved in superconducting accelerator magnet R&D under the framework of the U.S. Magnet Development Program [1]. An integral part of that program is High Temperature Superconducting (HTS) accelerator magnet development to demonstrate self-fields of 5 T or greater compatible with operation in hybrid configurations to generate fields beyond 16 T for future High Energy Physics (HEP) applications.

The Conductor on Molded Barrel (COMB) magnet technology [2] is being developed to address the ever-increasing requirements on the magnetic field strength from the physics community, which lead to high levels of mechanical stresses in the coils and degradation of conductor properties. A typical COMB coil consists of several layers of conductors wound into a contiguous channel without inner joints. Each turn is supported by the structure to control its position both in the straight section and the ends and to prevent the force accumulation between the turns – i.e. the so called "stress management." It is distinguished from other magnet designs employing the stress management, including the Canted Cosine-Theta (CCT) accelerator magnet concept, proposed in 2014 [3] and further developed for Nb$_3$Sn [4]-[5], REBCO [6] and Bi-2212 [7] materials in that the conductors are parallel to the magnet axis in the straight section. It also differs from the Stress Managed Cosine-Theta (SMCT) concept proposed in 2017 [8] and further developed for Nb$_3$Sn [9]-[10] in that the channel contains a single and typically round conductor.

The COMB design works well with Symmetric Tape Round (STAR®) wires produced by AMPeers LLC, which are among the most promising REBCO conductors for accelerator magnets [11]. Due to the proprietary architecture placing the superconducting layer near the neutral plane of the tape [12]-[13], they offer unrivaled bending performance suitable for accelerator magnets for future HEP experiments.

A two-layer REBCO-STAR dipole magnet with a 60 mm clear bore, named COMB-STAR-1 has been fabricated with about 10 m of STAR® wire and tested in liquid nitrogen in 2023, demonstrating over 90% of the critical current ($I_c$) retention in both coils measured before and after winding [14].

Even though the magnet was not originally designed for operation at a lower than 77 K temperature, it was decided to test it in liquid helium to: check the instrumentation and the data acquisition system; evaluate the performance of the magnet protection system on such a magnet; gather the information on possible magnet design changes and upgrades of the test facility for future testing of HTS magnets.

The magnet was re-assembled with a larger iron yoke and axial supports and tested in liquid helium in 2024 at Fermilab's Vertical Magnet Test Facility [15]. It was the first HTS magnet test at that facility.

## II. MAGNET DESIGN AND RE-ASSEMBLY

The COMB-STAR-1 magnet solid model and assembly are shown in Fig. 1, and Fig. 2 shows the magnetic field distribution in the coil and iron yoke. The magnet consisted of two half-coils, each wound from about 5 m of STAR® wire in a double-pancake configuration without the inner joints. The bare conductor was supported in the insulating COMB structures 3D printed from ULTEM™ 1010 with a nearly 100% fill factor. The leads of two half-coils were connected through a copper adapter. Two additional adapters attached the coils to NbTi cables spliced to the test facility leads.

This manuscript has been authored by Fermi Research Alliance, LLC under Contract No. DE-AC02-07CH11359 with the U.S. Department of Energy, Office of Science, Office of High Energy Physics, through the US MDP. This work was supported by the U.S. Department of Energy SBIR award DE-SC0022900.

Corresponding author: Vadim V. Kashikhin.

V. V. Kashikhin, S. Cohan, J. DiMarco, O. Kiemschies, S. Krave, V. Lombardo, V. Marinozzi, D. Orris, S. Stoynev and D. Turrioni are with Fermi National Accelerator Laboratory, Batavia, IL 60510 USA (e-mail: vadim@fnal.gov).

A. K. Chavda, S. Korupolu, J. Peram, A. Arjun, C. Goel, V. Yerraguravagari, R. Schmidt, V. Selvamanickam and G. Majkic are with Department of Mechanical Engineering, Advanced Manufacturing Institute, Texas Center for Superconductivity, University of Houston, Houston, TX 77204, USA (e-mail: selva@uh.edu). VS has financial interest in AMPeers.

E. Galstyan, N. Mai, U. Sambangi, J. Sai Sandra, and K. Selvamanickam are with AMPeers LLC, Houston, TX 77059, USA (e-mail: contact@ampeers-llc.com).

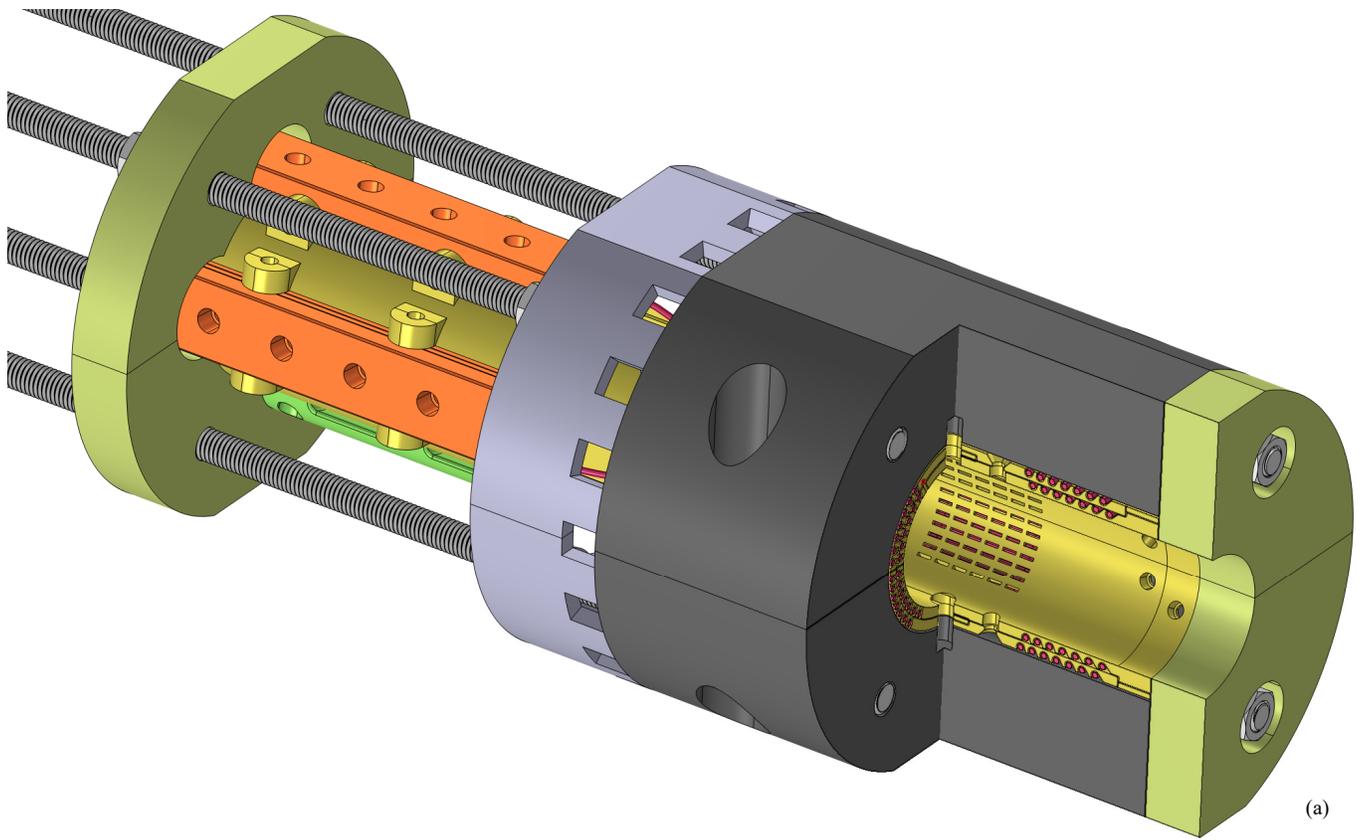

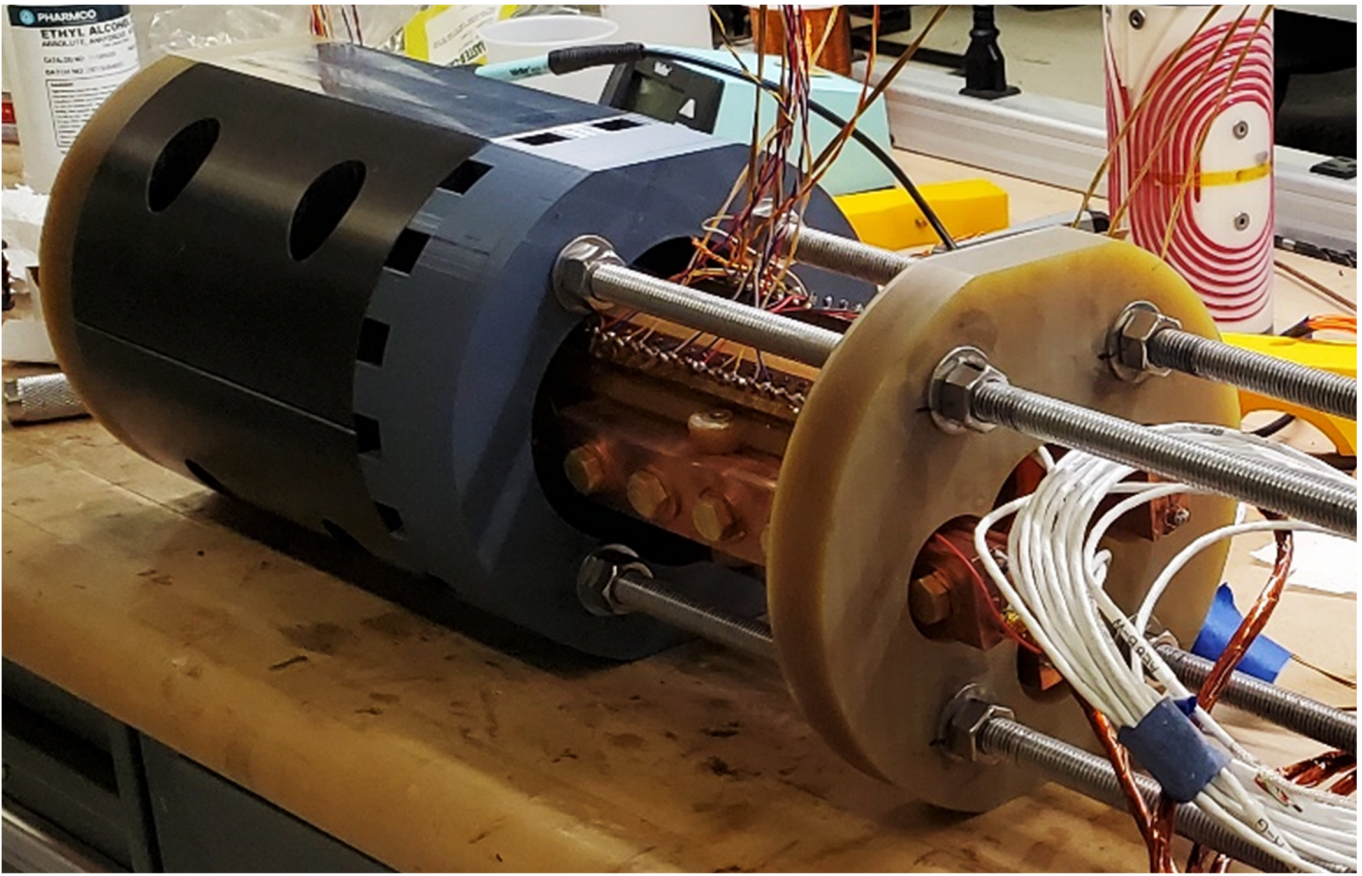

**Fig. 1.** The magnet solid model (a) - parts of the structure are suppressed for clarity, and the assembled magnet (b).

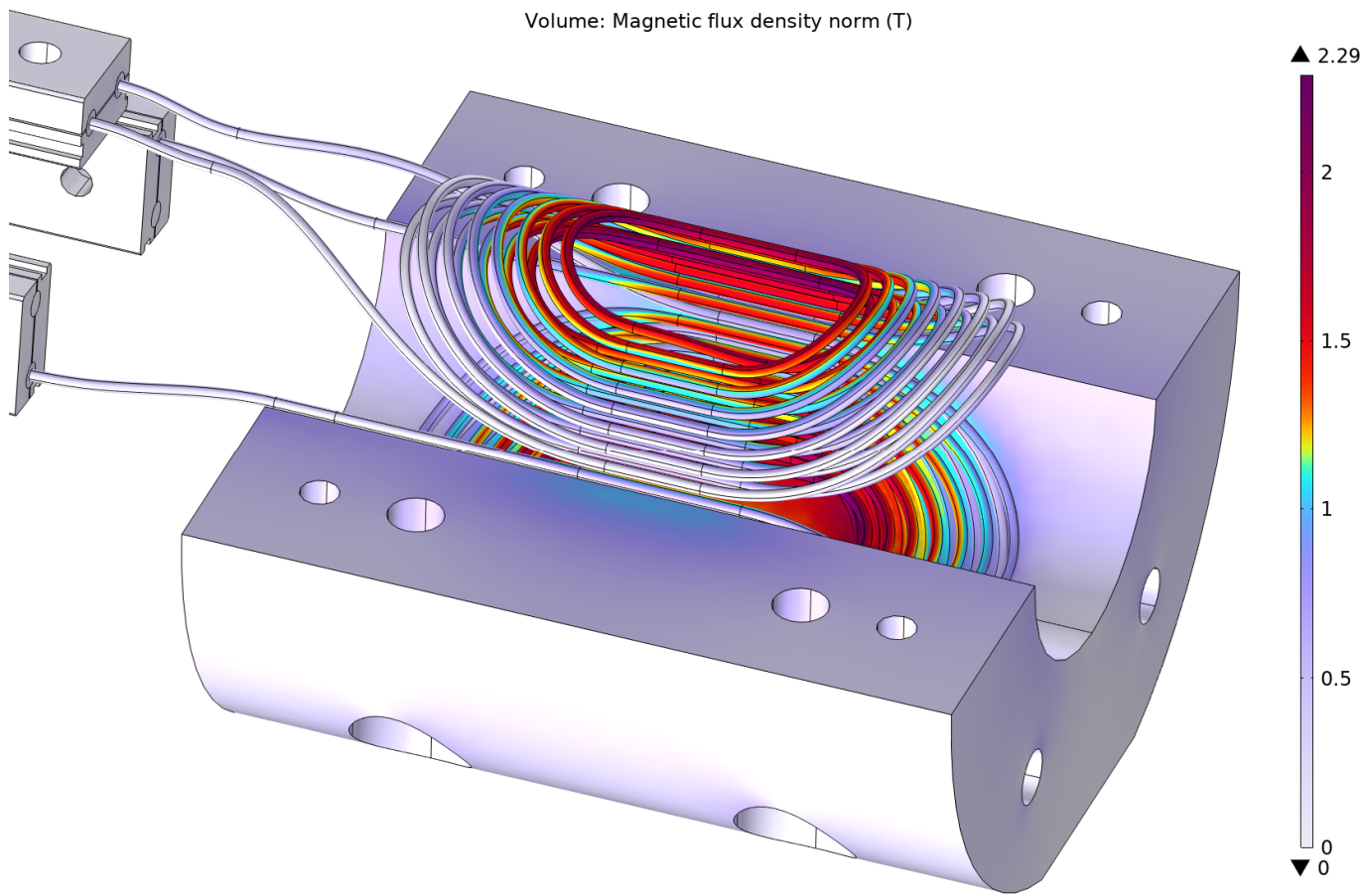

**Fig. 2.** Magnetic field in the coil and iron yoke at 3.3 kA current. Top half of the yoke is suppressed for clarity.

The iron yoke was fabricated from AISI 1018 steel. It consisted of two solid pieces with a split in the magnet midplane that was fully closed during the assembly. The coil structure was preloaded by means of Kapton shims between the coil and the iron yoke that created a radial interference of 0.25 mm. Since the conductor was wound relatively loosely into the channel, the main purpose of the shims was to avoid developing a gap between the coil and the yoke after the cooldown rather than preloading the conductor.

In the axial direction, the coil support structure was preloaded by means of two G-10 end plates and four Aluminum 6061-T6 rods. The axial interference was selected to compensate the differential thermal contraction between the aluminum rods and the ULTEM™ structure. The purpose of the axial preload was to create the rigid end boundaries to avoid overextending the structure with a relatively low elastic modulus of ~3 GPa under the Lorentz forces.

The main magnet parameters are listed in Table I. Since the coil was relatively short comparing to its diameter, there was not much difference between the peak field in the straight section and the ends as can be seen from Fig 2. Also, due to low iron field, the bore and peak coil field were linear vs. current and could be described by constant transfer functions.

The magnet was instrumented with redundant voltage taps co-wound with the conductor into the support structure and measuring the voltages across each half coil. In addition, there were eight acoustic gauges based on piezoelectric transducers, similar to those described in [16], attached to the inner surface of the coil and placed symmetrically ~20 degrees from the midplane at coil end. The magnet was also equipped with a dedicated quench antenna installed into the warm bore at the test facility. The details of the quench antenna design and the measurement results are discussed in a separate paper [17].

After the assembly, the magnet was Hi-Pot tested to 1 kV, applied between the conductor and the iron yoke with all the instrumentation wires floating electrically. The leakage current was under 0.1 μA.

TABLE I
COMB-STAR-1 MAGNET PARAMETERS

| Parameter | Unit | Value |
|---|---|---|
| Clear bore ID / yoke OD | mm | 60 / 190 |
| Coil and yoke length | mm | 190 |
| Total number of turns | - | 28 |
| Number of REBCO tapes per conductor | - | 11 |
| Width of inner / outer REBCO tapes | mm | 2.0 / 2.6 |
| Copper core OD | mm | 1.0 |
| Conductor OD | mm | 2.4 |
| Area of Cu / substrate / voids per conductor | mm$^2$ | 1.43 / 1.07 / 1.71 |
| Bore field / peak field transfer function | T/kA | 0.453 / 0.694 |
| Magnet inductance | μH | 72.6 |

## IV. Magnet Testing

The magnet was tested at 1.8 K – 4.5 K temperature at the magnet test facility. Fig. 3 shows the magnet attached to the top plate just prior to insertion in the vertical cryostat.

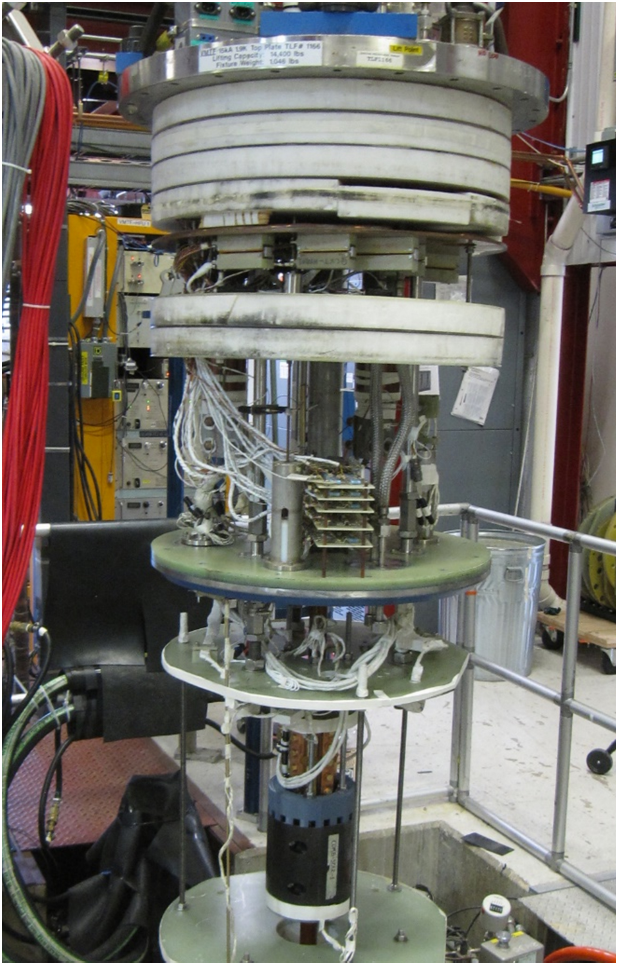

**Fig. 3** Magnet at the test facility.

During testing, the primary quench detection signal was the differential (bucked) voltage between the two half-coils. Since the half-coils were identical, it allowed to eliminate most (but not all) of the inductive noise pickup. The voltages were monitored by two separate systems based on Digital Quench Detection and Field Programmable Gate Arrays. Once either system detected the voltage above the threshold, it would trigger the quench protection to shut down the power supply and discharge the magnet through a dump resistor. The initial quench detection threshold was set at ±1 mV. In most cases, the magnet current was ramped up at a rate of 5-20 A/s.

The magnet quench history is shown in Fig. 4. The testing started at 1.8 K and the first 6 events were trips due to noise in the system. The quench detection threshold was progressively increased up to ±2.5 mV, at which point, the quench protection started to be triggered by the real resistive voltages. That threshold remained throughout the test, which was still a pretty small number comparing to typical quench detection voltages in low-temperature superconducting magnets.

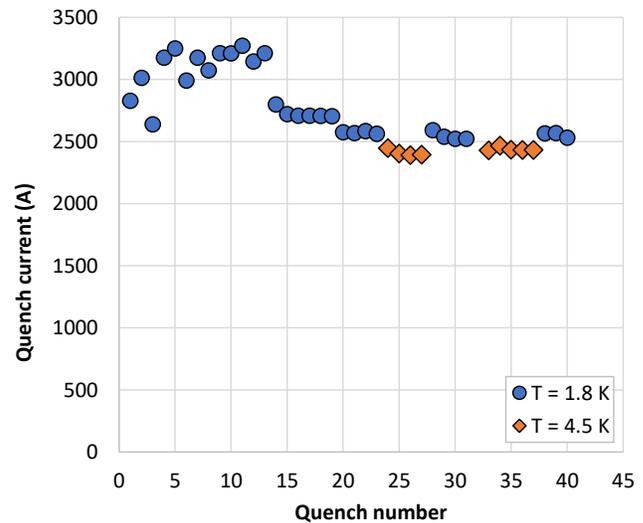

**Fig. 4** Magnet quench history.

During the next 7 (real) quenches, the maximum magnet current was relatively stable between 3100-3300 A with no apparent signs of training nor degradation. The highest current reached was 3273 A, which corresponded to 1.5 T bore field and 2.3 T peak coil field.

At that point, it was decided to continue with the rest of the testing program, which included precise Voltage-Current (*V-I*) measurements to capture the resistive transitions in the coils. The magnet current was ramped up in a stairstep pattern with 200 A steps and the *V-I* measurements were performed at each plateau. During the measurements at the 2800 A plateau, the magnet unexpectedly quenched. That event was no different from previous quenches at higher currents, however it likely degraded the conductor since all subsequent quenches were below that current value. As can be seen from the quench history, there were several other, albeit smaller drops in the quench current throughout the testing. Raising the temperature to 4.5 K resulted in lower quench currents, although, as expected, the difference with respect to 1.8 K was relatively small due to the high critical temperature of the conductor. All the quenches were localized in half-coil 2. The same conductor showed a small, but progressive reduction of $I_c$ when tested standalone (before winding into a coil) at 77 K [14]. It appears that whatever caused that reduction in liquid nitrogen was amplified when tested in liquid helium.

The typical voltage signals that triggered the quench protection are shown in Fig. 5. Even though the difference in the quench currents between 1.8 K and 4.5 K was small, the voltage signal behavior was drastically different. At 1.8 K, the quench development was characterized by a fast and mostly linear voltage growth that reached the threshold within ~35 ms from the quench onset. On the other hand, at 4.5 K, the quench development took much longer, with the slow linear voltage grows for ~350 ms. That difference is likely due to better cooling conditions in superfluid helium that allowed to ramp the current to higher values, which resulted in higher power dissipations and faster quench development after the heating power exceeded the cooling capacity.

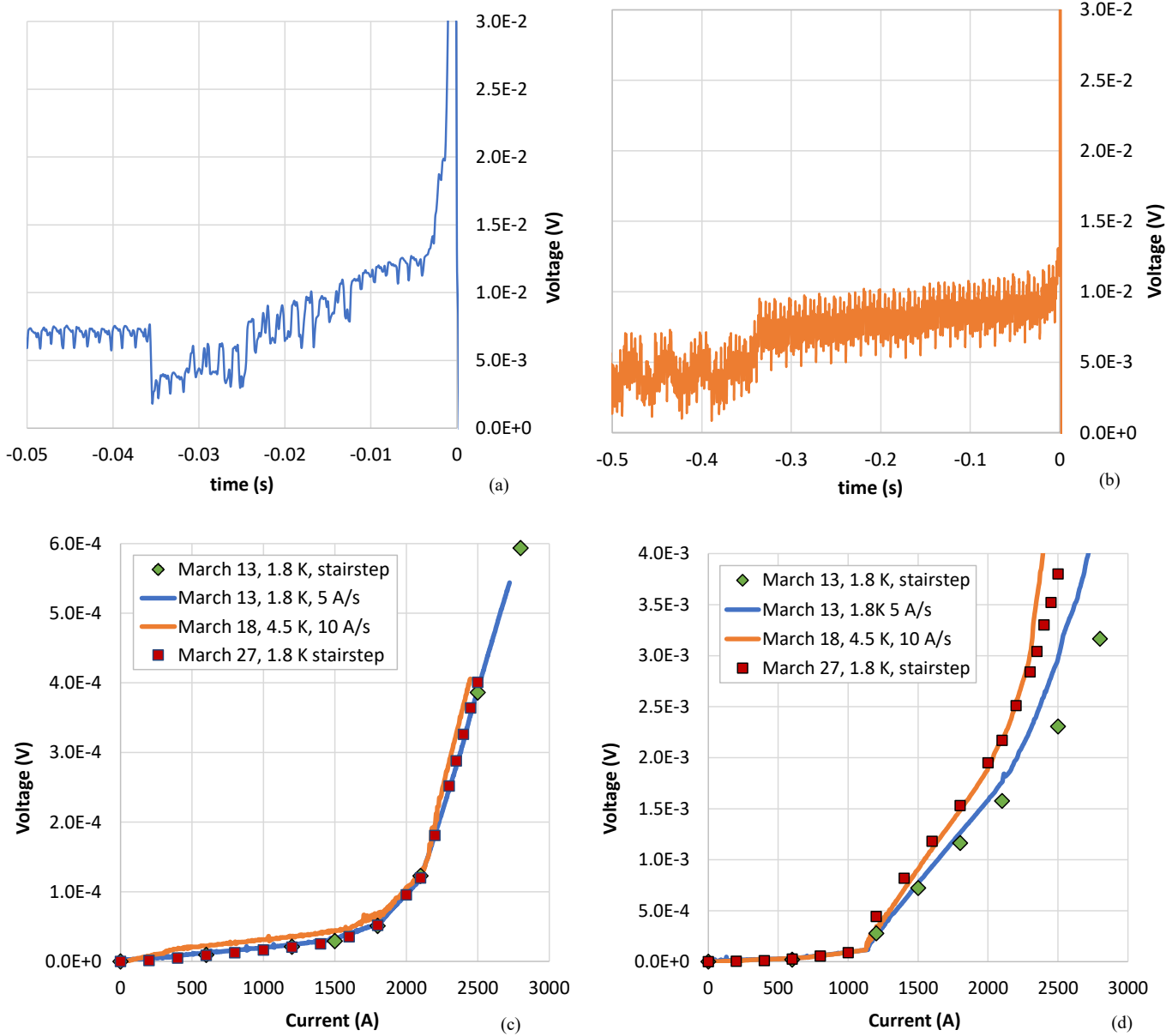

**Fig. 5.** Typical quench trigger signals at 1.8 K (a), 4.5 K (b) and the *V-I* measurements in half-coil 1 (c) and half-coil 2 (d).

Nevertheless, the last 5 ms before the quench detection in both cases was quite similar and characterized by a rapid acceleration of the voltage growth. It can be presumed that by this point all the liquid helium was either expelled from the channel by the building up pressure of entered the film boiling mode, which effectively stopped the cooling and further quench development happened under mostly adiabatic conditions. That characteristic delay time will be used to estimate the peak conductor temperature later in this section.

The *V-I* measurements were performed on different days during the testing with 2 weeks span before the first and the last test. Fig. 5 shows the *V-I* data in half-coils 1 and 2. The first and the last measurement in this set (solid markers) were the stairstep measurements as described above. The measurements in between were performed during current ramps up with a constant rate (solid lines). There was a slightly larger voltage during the ramps than at the plateaus due to the inductive voltage component. It is only mentioned to explain the visual difference in the plots; that effect was inconsequential for determining the $I_c$ since the linear voltage components (either resistive or inductive) were removed in subsequent analysis.

The voltage of half-coil 1 exhibited a linear growth until about 1.5 kA, which then turned into the typical power law behavior. For the half-coil 2, there was a drastic increase in the voltage slope (i.e. the constant resistance) at about 1.2 kA. The voltage remained linear vs. current up to 2 kA and only then followed the power law. Overall, the voltage of the half-coil 2 was a factor of 6-10 higher than that of the half-coil 1 at the same current values.

According to these data, the peak power dissipations in the half-coil 2 exceeded 10 W so it was obvious that the quench had a thermal rather than a mechanical onset.

The other important difference between the half coils in terms of the *V-I* performance comes from comparing the first and the last (stairstep) measurement. For the half-coil 1, the data mostly overlap with practically no visible difference. For the half-coil 2, there is a clear increase of the voltage from the beginning to the end of the testing campaign. It is consistent with the progressively reduced quench currents in Fig. 4.

After the test in liquid helium, the magnet was warmed up to the room temperature and then re-tested in liquid nitrogen. Table II shows the half-coil performance during these and previous tests.

To determine the $I_c$, the initial constant voltage slope was removed and then the $I_c$ was determined using the resistive criteria of 0.1 µV/cm. The n-value was derived from the least squares fit of the resistive power law to the measurement data. The half-coil 2 achieved 83% $I_c$ retention (the ratio between the last and the first *V-I* measurement) experiencing a noticeable degradation during the liquid helium tests. When re-tested in liquid nitrogen, it showed even lower $I_c$ retention of 60% compared to the previous test in liquid nitrogen. On the other hand, the half-coil 1 achieved 95% $I_c$ retention during the liquid helium tests. However, when re-tested in liquid nitrogen, it showed only 69% $I_c$ retention, indicating that it also degraded during the liquid helium testing.

TABLE II
COIL CRITICAL CURRENTS AND N-VALUES

| Condition | Temp (K) | Half-coil 1 | | Half-coil 2 | |
|---|---|---|---|---|---|
| | | $I_c$ (A) | n | $I_c$ (A) | n |
| Before LHe test | 77 | 443 | 6.4 | 422 | 7.5 |
| After LHe test | | 304 | 5.5 | 253 | 4.8 |
| $I_c$ retention | | 0.69 | | 0.60 | |
| First *V-I* measurement | 1.8 | 2366 | 7.3 | 1343 | 4.0 |
| Last *V-I* measurement | | 2245 | 7.3 | 1112 | 3.8 |
| $I_c$ retention | | 0.95 | | 0.83 | |
| Lift factor along load line | | 7.4 | | 4.4 | |

The magnet was protected by a dump resistor to extract the stored energy. The dump activation system required relatively large capacitors discharged across the Silicon Controlled Rectifiers (SCRs) in series with the power supply and the magnet to drive the current into the dump resistor [15]. Due to the low magnet inductance, it caused ~5 kA pulses on top of the transport current, as shown in Fig. 6, in all the events.

The effect of the test dynamics on the magnet temperature was analyzed using an adiabatic approach [18] assuming a 5 ms detection delay as mentioned earlier, followed by the current pulse on top of the 3.3 kA transport current and a subsequent current decay through the 60 mΩ dump resistor.

Fig. 7 shows the conductor capacity and the contributors to the MIITs budget. Under these conditions, the peak coil temperature would not exceed 60 K; however, this model assumed a uniform current distribution in the conductor, which might not be the case due to the fast 5 MA/s transients.

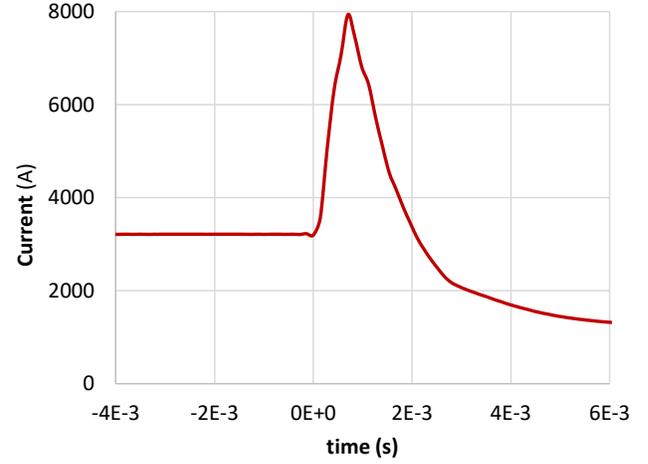

Fig. 6. A typical current pulse during testing.

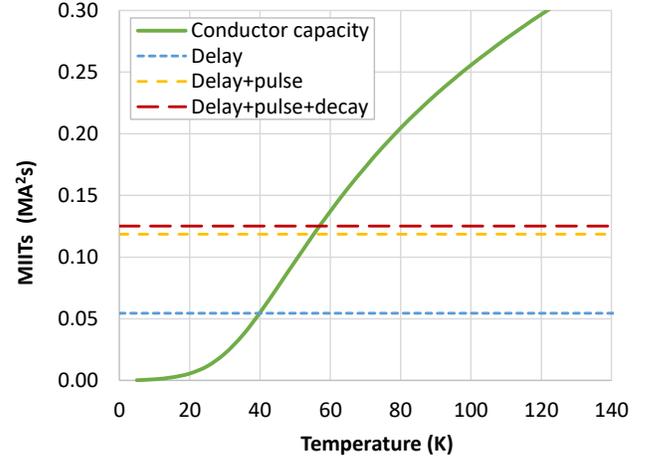

Fig. 7. Conductor capacity and MIITs budgets.

Since the tapes in the conductor were not fully transposed, the skin effect could drive a significant fraction of the current into the outer layers causing overheating and damage to the superconductor. Another issue is a considerable increase of the Lorentz forces during the current pulse that could further exacerbate the damage. A detailed analysis of these effects goes beyond the scope of this paper.

V. CONCLUSION

A REBCO dipole magnet based on STAR® wire from AMPeers LLC was fabricated and tested in liquid nitrogen in 2023 and liquid helium in 2024. The magnet reached 1.5 T field in a 60-mm bore and 2.3 T in the coil. The magnet performance was limited by one half-coil, but the critical currents of both half-coils were degraded during the liquid helium testing. The damage could have been caused by the test facility itself, which is being investigated.

On the other hand, the magnet survived 40 quenches, each one having pulses that exceeded the transport current by a large margin. Useful instrumentation data was collected that will be used for future magnets tests. It also indicated the necessity to upgrade the test facility to protect REBCO magnets, which is being implemented.

REFERENCES

[1] S. Prestemon, K. Amm, L. Cooley, S. Gourlay, D. Larbalestier, G. Velev and A. Zlobin, "A 2020 The 2020 Updated Roadmaps for the US Magnet Development Program," arXiv:2011.09539 [physics.acc-ph].

[2] V. V. Kashikhin, V. Lombardo and G. Velev, "Magnet Design Optimization for Future Hadron Colliders," *Proc. of 2019 Int. Part. Accel. Conf.*, THPTS084, doi: 10.18429/JACoW-IPAC2019-THPTS084.

[3] S. Caspi et al., "Canted-Cosine-Theta Magnet (CCT) – A Concept for High Field Accelerator Magnets", *IEEE Trans. Appl. Supercond.*, vol. 24, no. 3, June 2014, 4001804, doi:10.1109/TASC.2013.2284722.

[4] D. Arbelaez et al., "Status of the $Nb_3Sn$ Canted-Cosine-Theta Dipole Magnet Program at Lawrence Berkeley National Laboratory," *IEEE Trans. Appl. Supercond.*, vol. 32, no. 6, pp. 1-7, Sept. 2022, Art no. 4003207, doi: 10.1109/TASC.2022.3155505.

[5] B. Auchmann et al., "Test Results From CD1 Short CCT $Nb_3Sn$ Dipole Demonstrator and Considerations About CCT Technology for the FCC-Hh Main Dipole," *IEEE Trans. Appl. Supercond.*, vol. 34, no. 5, pp. 1-6, Aug. 2024, Art no. 4000906, doi: 10.1109/TASC.2023.3344425

[6] X. Wang et al., "Development and performance of a 2.9 Tesla dipole magnet using high-temperature superconducting CORC® wires," *2021 Supercond. Sci. Technol.*, 34 015012, doi: 10.1088/1361-6668/abc2a5.

[7] T. Shen et al., "Design, fabrication, and characterization of a high-field high-temperature superconducting Bi-2212 accelerator dipole magnet," *Phys. Rev. Accel. and Beams* 25, 122401 (2022), doi: 10.1103/PhysRevAccelBeams.25.122401.

[8] V. V. Kashikhin, I. Novitski and A.V. Zlobin, "Design Studies and Optimization of High-Field $Nb_3Sn$ Dipole Magnets for a Future Very High Energy PP Collider," *Proc. of 2017 Int. Part. Accel. Conf.*, WEPVA140, doi: 10.18429/JACoW-IPAC2017-WEPVA140.

[9] I. Novitski, A. V. Zlobin, E. Barzi and D. Turrioni, "Design and Assembly of a Large-Aperture $Nb_3Sn$ Cos-Theta Dipole Coil with Stress Management in Dipole Mirror Configuration," *IEEE Trans. Appl. Supercond.*, vol. 33, no. 5, pp. 1-5, Aug. 2023, Art no. 4001405, doi: 10.1109/TASC.2023.3244894.

[10] A.V. Zlobin, M. Baldini, I. Novitski, D. Turrioni, E. Barzi, "Development and test of a large-aperture $Nb_3Sn$ cos-theta dipole coil with stress management," *Proc. of 2024 Int. Part. Accel. Conf.*, WEPS68, doi: 10.18429/JACoW-IPAC2024-WEPS68.

[11] S. Kar, W. Luo, A. B. Yahia, X. Li, G. Majkic and V. Selvamanickam "Symmetric tape round REBCO wire with Je (4.2 K, 15 T) beyond 450 A/mm2 at 15 mm bend radius: a viable candidate for future compact accelerator magnet applications," *2018 Supercond. Sci. Technol.* vol 31, no. 4, 31, 04LT01, doi: 10.1088/1361-6668/aab293.

[12] S. Kar, J. Sai Sandra, W. Luo, M. Kochat, J. Jaroszynski, D. Abraimov, G. Majkic and V. Selvamanickam, "Next-generation highly flexible round REBCO STAR wires with over 580 A/mm2 at 4.2 K, 20 T for future compact magnets," *2019 Supercond. Sci. Technol.*, vol. 32, no. 10, 10LT01, doi 10.1088/1361-6668/ab3904.

[13] E. Galstyan, J. Kadiyala, M. Paidpilli, C. Goel, J. Sai Sandra, V. Yerraguravagari, G. Majkic, R. Jain, S. Chen, Y. Li, R. Schmidt, J. Jaroszynski, G. Bradford, D. Abraimov, X. Chaud, J. Song and V. Selvamanickam, "High Critical Current STAR® Wires with REBCO Tapes by Advanced MOCVD," *2023 Supercond. Sci. Technol.* vol. 36, no. 5, 055007, doi 10.1088/1361-6668/acc4ed.

[14] V. V. Kashikhin, S. Cohan, V. Lombardo, D. Turrioni, N. Mai, A. K. Chavda, U. Sambangi, S. Korupolu, J. Peram, A. Anil, C. Goel, J. Sai Sandra, V. Yerraguravagari, R. Schmidt, V. Selvamanickam, G. Majkic, E. Galstyan and K. Selvamanickam, "Accelerator magnet development based on COMB technology with STAR® wires," *2024 IOP Conf. Ser.: Mater. Sci. Eng.*, vol. 1301, 012153, doi: 10.1088/1757-899X/1301/1/012153.

[15] M. J. Lamm et al., "A new facility to test superconducting accelerator magnets," *Proc. of the 1997 Part. Accel. Conf.*, vol. 3, pp. 3395-3397, doi: 10.1109/PAC.1997.753220.

[16] M. Marchevsky, G. Sabbi, H. Bajas, S. Gourlay, "Acoustic emission during quench training of superconducting accelerator magnets," *Cryogenics*, vol 69, pp 50-57, 2015, ISSN 0011-2275, doi: 10.1016/j.cryogenics.2015.03.005.

[17] S. Stoynev, V. V. Kashikhin, S. Cohan, J. DiMarco, O. Kiemschies, S. Krave, N. Mai, U. Sambangi, V. Selvamanickam, "Application of Flex-QA Arrays in HTS Magnet Testing," *IEEE Trans. Appl. Supercond.*, to be published.

[18] M. N. Wilson, *Superconducting Magnets*, Oxford, UK: Oxford University Press, 1983, pp.201-202.